# Simultaneous electrical-field-effect modulation of both top and bottom Dirac surface states of epitaxial thin films of three-dimensional topological insulators

Cui-Zu Chang,[†, ‡, **] Zuocheng Zhang,[†, **] Kang Li,[‡, **] Xiao Feng,[†, ‡] Jinsong Zhang,[†] Minghua Guo,[†] Yang Feng,[†] Jing Wang,[∥] Li-Li Wang,[†, §] Xu-Cun Ma,[†, §] Xi Chen,[†, §] Yayu Wang,[†, §*] Ke He,[†, §*] and Qi-Kun Xue[†, §]

[†]State Key Laboratory of Low-Dimensional Quantum Physics, Department of Physics, Tsinghua University, Beijing 100084, China

[‡]Beijing National Laboratory for Condensed Matter Physics, Institute of Physics, Chinese Academy of Sciences, Beijing 100190, China

[∥]Department of Physics, McCullough Building, Stanford University, Stanford, CA 94305-4045, USA

[§]Collaborative Innovation Center of Quantum Matter, Beijing 100084, P. R. China.

**Abstract**: **It is crucial for the studies of the transport properties and quantum effects related to Dirac surface states of three-dimensional topological insulators (3D TIs) to be able to simultaneously tune the chemical potentials of both top and bottom surfaces of a 3D**

**TI thin film. We have realized this in molecular beam epitaxy-grown thin films of 3D TIs, as well as magnetic 3D TIs, by fabricating dual-gate structures on them. The films could be tuned between *n*-type and *p*-type by each gate alone. Combined application of two gates can reduce the carrier density of a TI film to a much lower level than with only one of them and enhance the film resistance by 10000 %, implying that Fermi level is tuned very close to the Dirac points of both top and bottom surface states without crossing any bulk band. The result promises applications of 3D TIs in field effect devices.**

A three-dimensional topological insulator (3D TI) is characterized by metallic surface states (SSs) resulting from the topological property of its insulating bulk bands.[1-4] With Dirac-cone-shaped energy band structure and spin-momentum locking property, topological SSs of 3D TIs are expected to show many novel physical properties and quantum phenomena which may be used to develop new concept electronic devices.[1-7] Ultra-high quality thin films of 3D TIs grown by molecular beam epitaxy (MBE) are favored in studies and applications of Dirac SSs for large surface/bulk conduction ratio, well-controlled thickness, and convenience of engineering their electronic structure.[8] MBE growth of $Bi_2Se_3$ family TIs, the most studied among TI materials, has been realized on many different substrates by several groups.[9-14]

Experimental observations of quantum phenomena predicted in a 3D TI thin film usually require electrical-field-effect structure fabricated on it, so that Fermi level can be fine-tuned into the bulk gap and only crossing the Dirac SSs.[1,2] There have been several reports on field-effect devices based on MBE-grown $Bi_2Se_3$ family TI thin films in either top- or bottom-gate geometry.[12,15-21] Single gate structure serves well for studies on usual semiconductors, graphene and two-dimensional (2D) TIs, whereas it is insufficient for 3D TI thin films. A 3D TI thin film has Dirac SSs at both its top and bottom surfaces, and the two surfaces may have different chemical potentials due to band-bending or different surface environments.[9] The transport data on such a film, including contributions from both surfaces, is difficult to interpret in straightforward way unless the chemical potentials of two surfaces are tuned nearly the same. It is particularly important for investigating the properties and quantum effects related to Dirac point, the most interesting part of a Dirac band, that we can simultaneously tune Dirac points of top and bottom SSs very close to Fermi level (see the top schematic in Figure 1a). On the other hand, some quantum phenomena, for example exciton condensation[22] and electrical field induced topological transition[23], are expected in a 3D TI thin film with its two surfaces tuned *p*- and *n*-doped, respectively (see the bottom schematic in **Figure 1a**). Dual-gate structure that can simultaneously tune the chemical potentials of two surfaces between *p*- and *n*-types should thus be a standard setup for studies of 3D TIs which however has rarely been applied in previous works on MBE-grown 3D TI thin films. In this study, by using $SrTiO_3$ substrate and amorphous $Al_2O_3$ layer prepared by atomic layer deposition (ALD) as bottom- and top-gate dielectrics, respectively, we realized simultaneous gate-tuning of the top and bottom SSs of three different 3D TI thin films between *p*- and *n*-types. This progress takes us a step closer to realizing many quantum effects and applications of 3D TIs.[1,2]

The samples used in this work are a 4QL $(Bi_{0.04}Sb_{0.96})_2Te_3$, a 4QL $Sb_2Te_3$ and a 5QL $Cr_{0.22}(Bi_{0.25}Sb_{0.85})_{1.78}Te_3$ films grown by MBE on $SrTiO_3$ (111) substrates. The film thicknesses chosen are large enough for these materials to develop gapless Dirac SSs.[24,25] The high dielectric constant of $SrTiO_3$ at low temperature (~20000 at 2K) makes the 0.25mm thick substrates still good dielectric layers for efficient bottom-gating.[12] **Figure 1b** shows the reflective high energy electron diffraction (RHEED) pattern taken during MBE growth of the 4QL $(Bi_{0.04}Sb_{0.96})_2Te_3$ film. Clear and sharp 1×1 diffraction streaks indicate 2D epitaxial growth and high crystalline quality of the film.[10] The surface band structures of the films have been checked by *in situ* angle resolved photoemission spectroscopy (ARPES). From the bandmap and corresponding momentum distribution curves of the 4QL $(Bi_{0.04}Sb_{0.96})_2Te_3$ film shown in **Figure 1c**, we can see characteristic linear band dispersion of Dirac SSs. The other two films studied in this work exhibit similar sample quality.

The dielectric layer for top-gate is amorphous $Al_2O_3$ of ~40nm prepared by ALD (Ensure NanoTech) through a two-step process (see **Figure 1d** top). An as-grown TI film is first covered with a 15nm thick $Al_2O_3$ by ALD with tri-methyl-aluminum (TMA) and $H_2O$ as precursors at temperature $T$=200°C, to protect the film from contamination during the following photolithography process. After being fabricated into Hall bar pattern with photolithography, the film is capped with another $Al_2O_3$ layer of 25nm by ALD with TMA and ozone as precursors at $T$=200°C. The second $Al_2O_3$ layer avoids electrical leakage at the edges of Hall bars between the below film and the above top-gate electrode layer (Ti/Au) evaporated later. A schematic diagram and an optical microscope image of a dual-gate Hall bar device are shown in **Figure 1d** bottom and **Figure 1e**, respectively. Transport measurements have been carried out at 300mK or 1.5K with standard ac lock-in technique.

First we tried 4QL $(Bi_{0.04}Sb_{0.96})_2Te_3$ film which is close to charge neutral and easy for gate-tuning.[26] **Figure 2a** displays the Hall trace (magnetic field ($\mu_0H$) dependence of the Hall resistance ($\rho_{yx}$)) of a Hall bar device made from the 4QL $(Bi_{0.04}Sb_{0.96})_2Te_3$ film with both gates grounded measured at 300mK. The Hall trace shows a straight line with positive slope indicating *p*-doping of the sample. The 2D carrier density ($n_{2D}$) is ~$1.3\times10^{13}$cm$^{-2}$ as estimated from the slope, i.e. Hall coefficient ($R_H$). The corresponding carrier mobility ($\mu$) is ~160cm$^2$Vs$^{-1}$. The carrier density value is higher than that in the film with the same composition grown on sapphire substrate,[26] probably caused by charge accumulation in SrTiO$_3$ substrate during sample cooling. The magnetoresistance (MR, magnetic field ($\mu_0H$) dependence of the longitudinal resistance ($\rho_{xx}$)) shown in the inset of **Figure 2a** exhibits a typical weak anti-localization behavior as usually seen in 3D TIs.[18]

By applying a bottom-gate voltage ($V_b$), we can tune the film from *p*-type to *n*-type. $V_b$ dependence of $R_H$ and $\rho_{xx}$ are displayed in **Figures 2b and 2c**, respectively. With increasing $V_b$, $R_H$ first increases, then rapidly drops from positive to negative, and resumes increasing after crossing a negative maximum. The $\rho_{xx}$-$V_b$ curve on the other hand only shows one maximum. These observations indicate a typical ambipolar behavior which means that dominating carriers in the film are tuned from holes to electrons by bottom-gating.[27] The whole $n_{2D}$ modulated by bottom-gating is ~2×10$^{13}$cm$^{-2}$ as $V_b$ changes from -10V to 100V (see the inset of **Figure 2c**).

Although ambipolar behavior is observed, we cannot obtain a very low carrier density. The minimum $n_{2D}$ measured is ~4.6×10$^{12}$cm$^{-2}$ (*n*-type). It is either because of energy difference between two the Dirac points, or of existence of other states overlapping the Dirac points. The former problem can be removed by applying both top- and bottom-gating. **Figures 2d and 2e** show top-gate voltage ($V_t$) dependence of $R_H$ and $n_{2D}$, respectively, measured at different $V_b$. All the data exhibit ambipolar behavior, and top-gating alone can change $n_{2D}$ of the film by several 10$^{13}$cm$^{-2}$. With increasing $V_b$, the charge neutral point ($R_H$=0) moves to lower $V_t$, indicating that the carriers introduced by one gate are compensated by the other. The observation suggests that we can separately tune the doping levels near the two surfaces of a TI film. For example, when $V_b$ = -10V and $V_t$ = 7V, the bottom surface is charged with holes whereas the top surface is charged with nearly the same number of electrons. The situation is reversed for $V_b$ = 210V and $V_t$ = -10V when the bottom and top surfaces are hole- and electron-doped, respectively. So we can engineer the potential profile across a TI thin film for realization of various quantum phenomena and applications by using top and bottom-gating together.

From **Figure 2f**, we notice that $V_t$ for maximum $\rho_{xx}$ is dependent on $V_b$, which suggests that there is coupling between top and bottom-gate voltages in the film. It is reasonable since the depletion length of the gates is around 30~35nm, estimated from the approximated expression $z_d = \sqrt{2k\varepsilon_0 \Delta E/n_{3D}e^2}$,[28] here $k$ is the dc dielectric constant of $Sb_2Te_3$ (~168),[29] $\varepsilon_0$ is the vacuum permittivity, $\Delta E$ is the band-bending energy by applying gate, and $n_{3D}$ is the 3D carrier density.

However the maximum $|R_H|$ of the $R_H$-$V_t$ curves is only slightly dependent on $V_b$ (see the dashed enveloping lines in **Figure 2d**). At $V_b$ = 35V and $V_t$ = 1V, $|R_H|$ reaches the maximum for all the $V_b$ and $V_t$ values, corresponding to $n_{2D}$ of $4.4 \times 10^{12}$cm$^{-2}$ (*n*-type). The failure in getting lower carrier density with dual-gate tuning can be attributed to the band structure of the material $(Bi_{0.04}Sb_{0.96})_2Te_3$. According to first principle calculations, the Dirac point of $Sb_2Te_3$ is very close to valence band maximum (VBM), and a small amount of Bi doped into $Sb_2Te_3$ can shift the Dirac point from bulk gap to below VBM.[18] In this case, the expected low carrier density when Fermi level crosses Dirac points can never been reached because of existence of bulk electrons. Pure $Sb_2Te_3$ still has its Dirac point exposed in bulk gap and can show low carrier density with dual-gate modulation. However $Sb_2Te_3$ is usually heavily hole-doped by anitsite defects. By optimizing MBE growth condition,[30] we managed to reduce the intrinsic carrier density of $Sb_2Te_3$ films to the level of ~$1 \times 10^{13}$cm$^{-2}$, low enough to be removed by gate voltages. **Figure 3a** shows the Hall traces of a 4QL $Sb_2Te_3$ film at different $V_t$ with $V_b$ = 0V. The Hall trace for $V_t$ = 0V gives a hole density of ~9.6×10$^{12}$cm$^{-2}$. By increasing $V_t$, the slope is significantly enhanced, indicating efficient reduction of carrier density. At $V_t$ = 7V, the hole density reaches ~ 1×10$^{12}$cm$^{-2}$, and at $V_t$ = 10V, the film becomes n-type with electron density of ~ 8×10$^{11}$cm$^{-2}$. The carrier mobility is ~320cm$^2$Vs$^{-1}$. By setting $V_t$ between 7V and 10V, one should get lower carrier density values. However sample resistance becomes so large that Hall signal becomes too noisy to be trusted. Therefore we can only evaluate the performance of dual-gating with $\rho_{xx}$.

In **Figure 3b** we show $\rho_{xx}$-$V_t$ curves of the 4QL Sb$_2$Te$_3$ film measured at different $V_b$. Dramatic changes induced by both top and bottom-gate voltages are observed. In the $\rho_{xx}$-$V_t$ curve for $V_b$ = 0V, $\rho_{xx}$ exhibits a peak up to 110kΩ at $V_t$ = 8V, about 40 times larger than that at $V_t$ = 0V. Assuming that carrier mobility of Dirac surface states is constant, the $n_{2D}$ corresponding to the $\rho_{xx}$ peak ($V_b$ = 0V, $V_t$ = 8V) should be ~2×10$^{11}$cm$^{-2}$ as estimated from the $n_{2D}$ and $\rho_{xx}$ values obtained at $V_b$ = 0V, $V_t$ = 10V. $\rho_{xx}$ is further enhanced by turning up bottom-gate voltage and reaches 240kΩ when $V_b$ = 60V and $V_t$ = 0.15V. When $V_b$ and $V_t$ are set far from the maximum, $\rho_{xx}$ is reduced to the order of 2kΩ, comparable with $\rho_{xx}$ of the 4QL (Bi$_{0.04}$Sb$_{0.96}$)$_2$Te$_3$ film with similar $n_{2D}$. So the high $\rho_{xx}$ can only be attributed to the low carrier density when Fermi level cuts two Dirac points simultaneously. We can observe a shoulder on the left side of some curves as indicated with dashed circles. It can be attributed to the edge of bulk valence bands which induce discontinuity in $\rho_{xx}$ when crossing Fermi level.[31]

According to previous scanning tunneling spectroscopy study, a 4QL thick Sb$_2$Te$_3$ film shows sharp zero-mode Landau level in magnetic field, which guarantees formation of gapless Dirac point at that thickness.[25] A 3QL thick Sb$_2$Te$_3$ film, on the other hand, exhibits a SS gap of

~50meV[25] and shows the maximum $\rho_{xx}$ above 1MΩ even when modulated with a single gate[30]. The much smaller $\rho_{xx}$ peak value of 4QL film (240kΩ) than 3QL one is consistent with the existence of gapless SSs. In single layer graphene, another well-known Dirac system, $\rho_{xx}$ usually shows a peak of around 6kΩ as Fermi level sweeps Dirac point.[31] One may concern why the maximum $\rho_{xx}$ in $Sb_2Te_3$ is much larger than that in graphene. The difference probably comes from different properties of the Dirac states of graphene and TIs. The Dirac point of graphene is protected by valley symmetry and, under zero magnetic field, can only be gapped by intervalley scattering which is usually quite weak.[33] The Dirac point of topological surface states of TIs is protected by time-reversal symmetry. The spin degeneracy of Dirac point of TIs is immune to elastic scattering but not necessary to inelastic one, or dephasing.[34] The latter process can only be neglected when sample size is smaller than the phase relaxation length ($L_\phi$), typically several hundred nanometers for $Bi_2Se_3$ family TIs.[35] In HgTe quantum well 2D TI that shows quantum spin Hall effect, $\rho_{xx}$ reaches ~100kΩ when sample size is increased to ~20μm because of inelastic scattering between two time-reversal edge channels. Strong backscattering near Dirac points of 3D TIs induced by dephasing together with impurity scattering has been demonstrated by theoretical and spectroscopic studies.[36,37] The sample studied here is much larger than $L_\phi$. It is thus reasonable that inelastic scattering leads to strong backscattering at Dirac point and a rather large maximum $\rho_{xx}$. In smaller dual-gated $Sb_2Te_3$ samples and at lower temperature where inelastic scattering is suppressed, the intrinsic transport properties of Dirac points can be revealed. It will be valuable to the TI field. Furthermore, the observed resistance enhancement by a factor of 10000% via gate-tuning, something not seen before, may find applications as field-effect transistors.

Dual-gate structure is also expected to help magnetic 3D TI thin films to show the quantum anomalous Hall effect which requires that the Fermi level simultaneously cross the magnetically induced gaps, typically only several meV wide, at both top and bottom SSs. **Figures 4a** shows the $\rho_{yx}$-$\mu_0 H$ curves of a dual-gated 5QL film of $Cr_{0.22}(Bi_{0.15}Sb_{0.85})_{1.78}Te_3$ measured at $T = 300$mK with different $V_b$ and $V_t$. The square-shaped hysteresis loops of anomalous Hall effect reflect good long-range ferromagnetism of the film. The inset in **Figure 4a** shows typical MR of the sample exhibiting a butterfly shape as commonly observed in ferromagnetic materials. The anomalous Hall resistance ($R_{AH}$, i.e. $\rho_{yx}$ at zero magnetic field) changes significantly with $V_b$ and $V_t$. $R_{AH}$ is only 400Ω with both $V_b$ and $V_t$ set 0V. By tuning $V_t$ to 15V, $R_{AH}$ increases to 1.8kΩ. When $V_b = 170$V and $V_t = +3.5$V, a maximum $R_{AH}$ of ~ 4.8kΩ is reached. **Figure 4b** displays $R_{AH}$-$V_t$ curves taken at different $V_b$. It is clear that higher $R_{AH}$ can be reached with both gate-voltages applied than only with one. However the maximum $R_{AH}$ measured is far below quantized value. We found that the process of ALD significantly reduces $R_{AH}$. Further attempts in top-gate fabrication procedure are needed to get dual-gate structure compatible with quantum anomalous Hall samples.[38]

In conclusion, we have realized high efficient dual-gate modulation in different 3D TI thin films grown by MBE. The films can be tuned between *n*- and *p*-types with each of the two gates alone, enabling controlling the potential profile of the films with large freedom. The Dirac points of both top and bottom SSs can be shifted simultaneously crossing Fermi level, which not only facilitates studies on intrinsic properties of Dirac point but also helps for the observation of the quantum anomalous Hall effect in magnetic TIs.[38] The obtained factor of 10000% change in On/Off ratio in $Sb_2Te_3$ promises applications of 3D TIs in field-effect transistors.

## ASSOCIATED CONTENT

Supporting Information. Detailed theoretical explanations for the hump structure in **Figure 3b** and gate dependence mobility of the films**.** This material is available free of charge via the Internet at http://pubs.acs.org.

## AUTHOR INFORMATION

*Corresponding authors. Email: yayuwang@tsinghua.edu.cn (Y.W.); kehe@tsinghua.edu.cn (K.H.)

**These authors contributed equally to this work.

**Funding Sources**

This work is supported by the National Basic Research Program of China (Grant Nos. 2013CB921702 and 2012CB921300) and the National Natural Science Foundation of China (Grant Nos. 11174343 and 11325421).

**Notes**

The authors declare no competing financial interest.

**Acknowledgment**

We wish to thank J. S. Moodera for helpful discussions, and J. Chen, Y. Q Li, W. X. Li and C. Z. Gu for technical support.

24. Li, Y. Y.; Wang, G.; Zhu, X.; Liu, M; Ye, C.; Chen, X.; Wang, Y. Y.; He, K.; Wang, L.; Ma, X.; Zhang, H. J.; Dai, X.; Fang, Z.; Xie, X.; Liu, Y.; Qi, X.; Jia, J. F. ; Zhang, S. C.; Xue, Q-K. *Adv. Mater.* **2010**, *22*, 4002–4007.

25. Jiang, Y. P.; Wang, Y. L.; Chen, M.; Li, Z.; Song,C. L.; He, K.; Wang, L. L.; Chen, X.; Ma, X. C. and Xue, Q. K.. *Phys. Rev. Lett.* **2012,** *108,* 016401-5.

26. Zhang, J.; Chang, C.Z.; Zhang Z.; Wen J.; Feng X.; Li K.; Liu M.; He K.; Wang, L.; Chen, X.; Xue, Q.; Ma X.; Wang Y. *Nat. Commun.* **2011**, *2*, 574-6.

27. Novoselov, K. S.; Geim, A. K.; Morozov, S. V.; Jiang, D.; Zhang,Y.; Dubonos, S. V.; Grigorieva, I. V.; Firsov, A. A. *Science* **2004**, *306*, 666–669.

28. Monch, W. *Semiconductor Surfaces and Interfaces*, Springer: New York, **2001**.

29. Madelung, O.; Rosser, U.; Schulz, M., *Non-Tetrahedrally Bonded Elements and Binary Compounds I*, Springer: pp.1–4, **1998**.

30. Jiang, Y. P.; Sun, Y. Y.; Chen, M.; Wang, Y. L.; Li, Z.; Song, C. L.; He, K.; Wang, L. L.; Chen, X.; Xue, Q. K.; Ma, X. C.; Zhang, S. B. *Phys. Rev. Lett.* **2012,** *108,* 066809-5.

31. See Supporting Information.

32. Private communication.

33. Sarma, S. D.; Adam, S.; Hwang, E. H.; Rossi, E.  *Rev. Mod. Phys.* **2011**, *83*, 407-470

34. König, M.; Wiedmann, S.; Brüne, C.; Roth, A. ; Buhmann, H.; Molenkamp, L.W.; Qi, X.; Zhang, S.-C. *Science* **2007**, *318*, 766–770.
Page **13** of **19**

**Figures and Captions:**

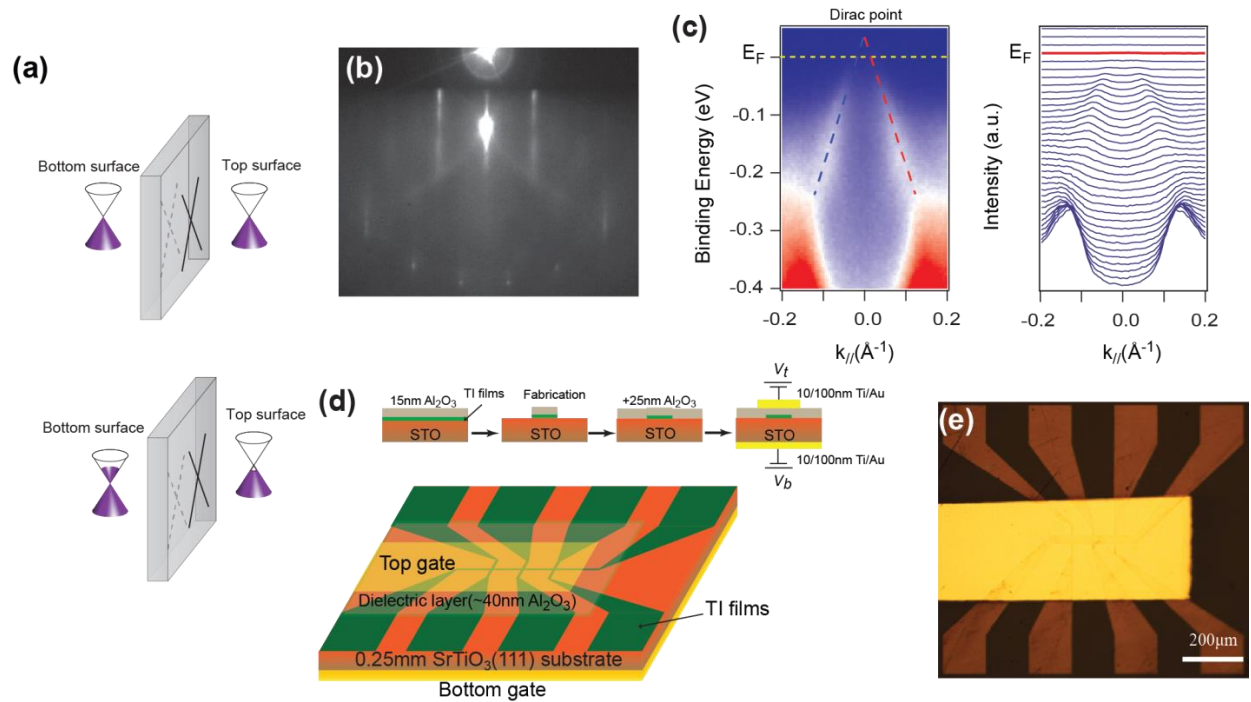

**Figure 1. (a)** Schematic diagrams of a 3D TI thin film with the Dirac points of both its top and bottom surfaces across Fermi level (top) and with its top and bottom surfaces hole- and electron-doped, respectively. **(b)** RHEED pattern of 4QL thick $(Bi_{0.04}Sb_{0.96})_2Te_3$ film. **(c)** ARPES band map and corresponding momentum distribution curves (MDCs) of 4QL $(Bi_{0.04}Sb_{0.96})_2Te_3$ film. **(d)** Schematic diagrams of the fabrication process (top) and structure (bottom) of dual-gate devices. **(e)** Optical micrograph of a dual-gate device.

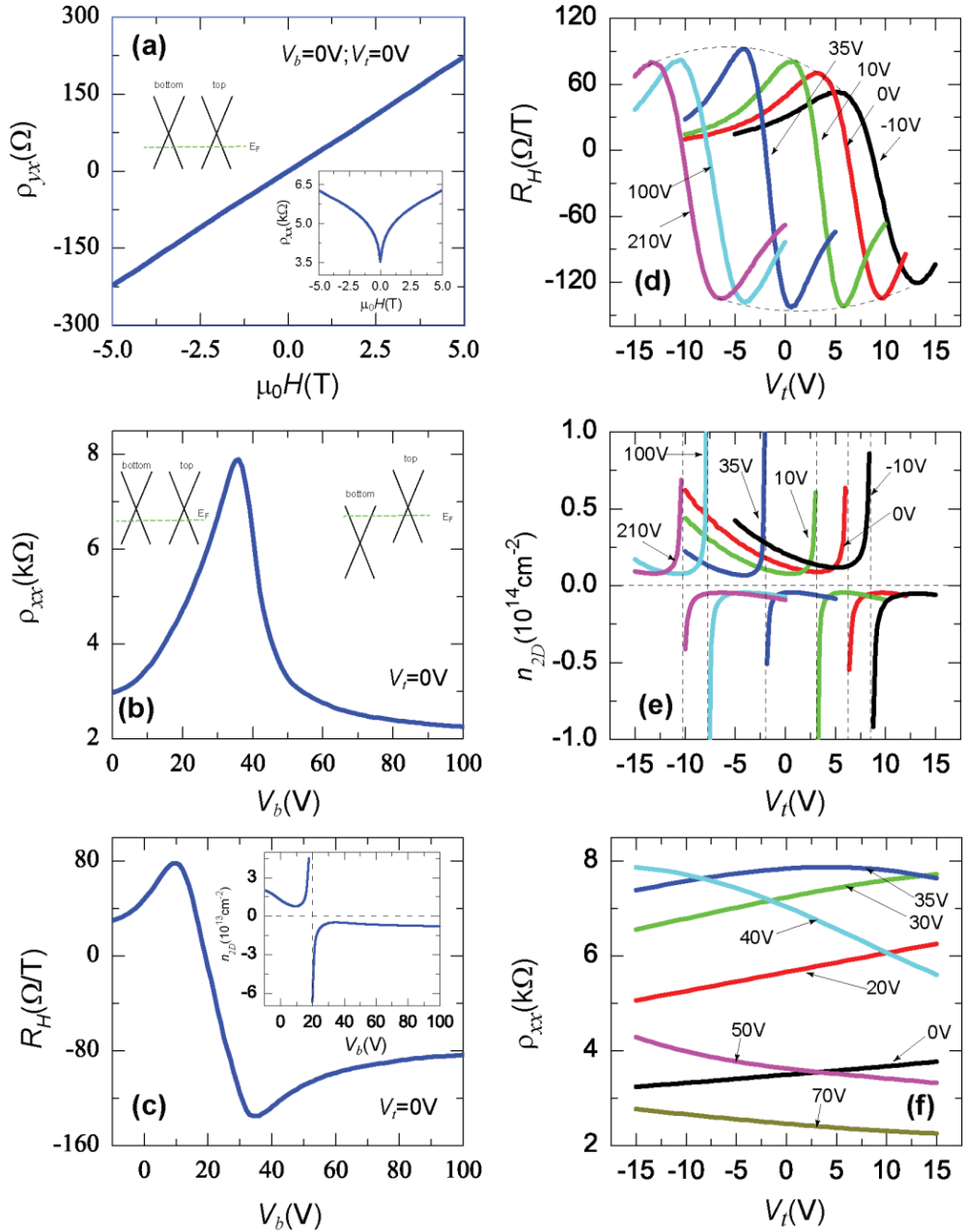

**Figure 2.** Transport results of 4QL $(Bi_{0.04}Sb_{0.96})_2Te_3$ film measured at 300 mK. **(a)** Magnetic field ($\mu_0H$) dependence of Hall resistance ($\rho_{yx}$). The inset shows $\mu_0H$ dependence of longitudinal resistance ($\rho_{xx}$). **(b,c)** Bottom gate voltage ($V_b$) dependence of $\rho_{xx}$ **(b)** and Hall coefficient ($R_H$) **(c)** with top gate voltage ($V_t$) set as 0V. The inset of **(c)** shows $V_b$ dependence of the nominal $n_{2D}$. **(d-f)** Top gate voltage ($V_t$) dependence of $R_H$ **(d)**, $n_{2D}$ **(e)**, and $\rho_{xx}$ **(f)** at different $V_b$. The sketch

diagrams in **(a)** and **(b)** identify the Fermi level position in the both top and bottom surfaces via tuning the gate bias.

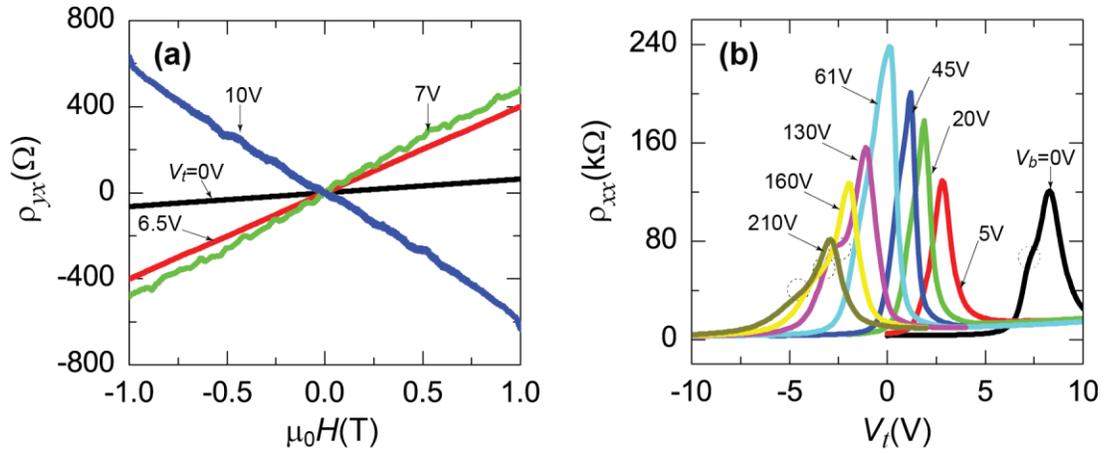

**Figure 3.** Transport results of 4QL $Sb_2Te_3$ film measured at 1.5 K. **(a)** $\mu_0 H$ dependence of $\rho_{yx}$ at different $V_t$ with $V_b$=0V. **(b)** $V_t$ dependence of $\rho_{xx}$ at different $V_b$.

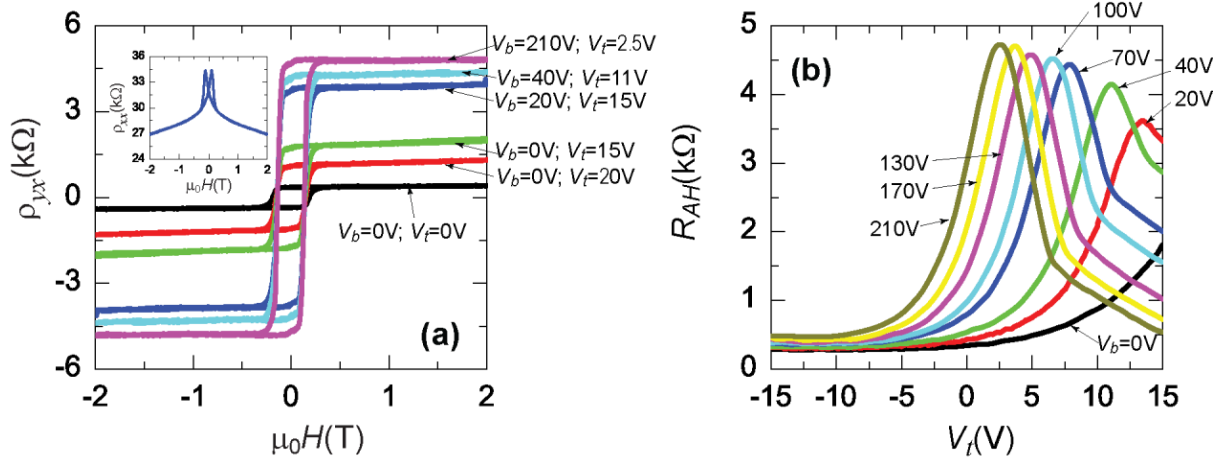

**Figure 4.** Transport results of 5QL $Cr_{0.22}(Bi_{0.25}Sb_{0.85})_{1.78}Te_3$ film measured at 300mK. **(a)** $\mu_0H$ dependence of $\rho_{yx}$ at different $V_b$ and $V_t$. The inset shows a $\rho_{xx}$-$\mu_0H$ curve of the sample. **(b)** $V_t$ dependence of anomalous Hall resistance ($R_{AH}$) at different $V_b$.

**TOC Graphic:**

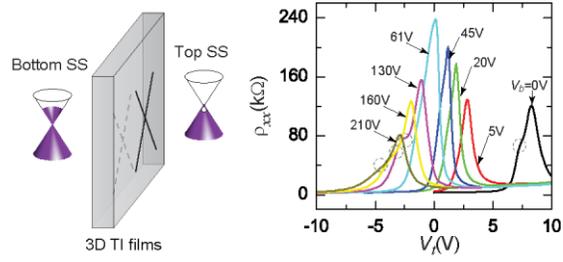